\documentclass[aps,nofootinbib,superscriptaddress,twocolumn,floatfix]{revtex4}
\usepackage{graphicx}
\usepackage{amsmath}
\usepackage{dcolumn}

\newcommand{\ChemF}{M$_{1-2x}$Co$_{1+x}$[Fe(CN)$_6$]$\cdot z$H$_2$O}
\newcommand{\musr}{$\mu$SR}
\newcommand{\Co}{Co$_{1.5}$}
\newcommand{\Cofull}{Co$_{1.5}$[Fe(CN)$_6$]$\cdot 6$H$_2$O}
\newcommand{\K}{K$_{0.2}$Co$_{1.4}$}
\newcommand{\Kfull}{K$_{0.2}$Co$_{1.4}$[Fe(CN)$_6$]$\cdot 6.9$H$_2$O}
\newcommand{\Rb}{Rb$_{0.66}$Co$_{1.25}$}
\newcommand{\Rbfull}{Rb$_{0.66}$Co$_{1.25}$[Fe(CN)$_6$]$\cdot 4.3$H$_2$O}

\begin{document}

\title{Muon Spin Relaxation Study of the Magnetism in Unilluminated Prussian Blue Analogue Photo-magnets}
\author{Z.~Salman}
\affiliation{TRIUMF, 4004 Wesbrook Mall, Vancouver, BC, Canada, V6T 2A3}
\author{T.J. Parolin} 
\affiliation{Chemistry Department, University of British Columbia, Vancouver, BC, Canada V6T 1Z1}
\author{K.H.~Chow}
\affiliation{Department of Physics, University of Alberta, Edmonton, AB, Canada T6G 2J1}
\author{T.A.~Keeler}
\affiliation{Department of Physics and Astronomy, University of British Columbia, Vancouver, BC, Canada V6T 1Z1}
\author{R.I.~Miller}
\affiliation{TRIUMF, 4004 Wesbrook Mall, Vancouver, BC, Canada, V6T 2A3}
\author{D.~Wang}
\affiliation{Department of Physics and Astronomy, University of British Columbia, Vancouver, BC, Canada V6T 1Z1}
\author{W.A.~MacFarlane}
\affiliation{Chemistry Department, University of British Columbia, Vancouver, BC, Canada V6T 1Z1}

\begin{abstract}
  We present longitudinal field muon spin relaxation ($\mu$SR)
  measurements in the unilluminated state of the photo-sensitive
  molecular magnetic Co-Fe Prussian blue analogues \ChemF, where M=K
  and Rb with $x=0.4$ and $\simeq 0.17$, respectively. These results
  are compared to those obtained in the $x=0.5$ stoichiometric limit,
  Co$_{1.5}$[Fe(CN)$_6$]$\cdot 6$ H$_2$O, which is not
  photo-sensitive. We find evidence for correlation between the range
  of magnetic ordering and the value of $x$ in the unilluminated state
  which can be explained using a site percolation model.
\end{abstract}

\pacs{75.50.Lk,76.75.+i,75.50.Xx,78.90.+t}

\maketitle

\subsection{Introduction}
Prussian Blue (PB) (Fe$_4$[Fe(CN)$_6$]$_3$) is a long-known dye and
prototypical transition metal co-ordination compound that exhibits
ferro-magnetism \cite{Holden56PR,Bozorth56ibid,Herren80IC} driven by
superexchange coupling between iron spins. It is an important case of
exchange coupling mediated through the CN$^-$ bridge
\cite{Ginsberg71ICAR}. Much of the recent interest in magnetic
compounds related to PB (including a few that have been studied with
$\mu$SR \cite{Salman02PRB,Salman00PB,Jestadt01JPCM}) is motivated by
potential novel behavior and related applications in molecular
magnetism \cite{ItohKinoshita,Sato96S}, including magnetism that is
sensitive to exposure to light, i.e. {\it photomagnetism}.

\begin{figure}
\includegraphics[width=\columnwidth]{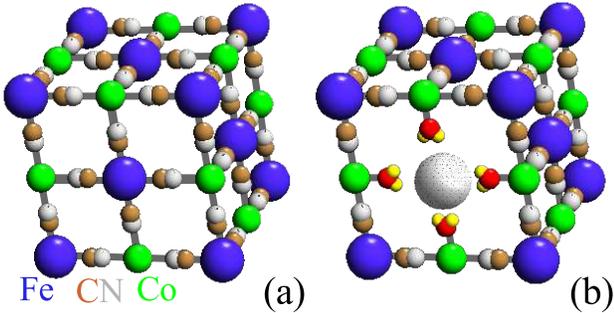}
\caption{(Color online) Crystal structure of \ChemF: (a) $x=0.5$; (b)
  $x \ne 0.5$, with an Fe(CN)$_6$ vacancy shown in gray coordinated by
  bound water molecules. Large, medium and small circles denote Fe,
  Co, and CN, respectively. The alkali ions which maintain charge
  neutrality occupy cubic interstitial sites (not shown).}
\label{structure}
\end{figure}

The compounds studied in this paper are the molecule based Co-Fe PB
analogues (Co-Fe PBAs) \ChemF\footnote{This chemical formula is a
  commonly used approximation though the ratio M:Co may be slightly
  different \cite{Pejakovic00JAP,Shimamoto02IC}.} (M is an alkali
metal) \cite{Sato96S,ItohKinoshita,Pejakovic01SM}. These compounds
have the sodium chloride structure, with Co and Fe ions located on the
vertices of a cubic lattice, each octahedrally coordinated by six
cyano moieties (Fig.~\ref{structure}).
The Co and Fe ions are connected via cyanide bridges with interstitial
alkali metal ions and water molecules \cite{Sato96S,Sato99IC}
(Fig.~\ref{structure}(b)). Co-Fe PBAs are in general
non-stoichiometric and significant structural disorder (vacancies in
the Fe(CN)$_6$ sites) is present. Depending on the stoichiometry and
synthesis route, the materials are paramagnets or exhibit magnetic
ordering at temperatures below $\sim 25$ K, due to small superexchange
coupling $J$ between Fe$^{III}$ and Co$^{II}$ moments.

Illumination of Co-Fe PBAs with broadband visible light in the range
of $\sim 550-750$ nm can cause dramatic changes in the magnetic
properties, including an increase in magnetization and ordering
temperature. The proposed origin of the photomagnetic effect is a
light-induced charge transfer from the state:
Fe$^{II}(t_{2g}^6,S=0)$-CN-Co$^{III}(t_{2g}^6,S=0)$ to the meta-stable
self--trapped state:
Fe$^{III}(t_{2g}^5,S=1/2)$-CN-Co$^{II}(t_{2g}^5e_g^2,S=3/2)$
\cite{Sato96S,Kawamoto01PRL}, effectively increasing the concentration
of magnetic moments. The non-stoichiometry is believed essential for
the photoinduced magnetization \cite{Kawamoto01PRL}. However, in spite
of considerable experimental and theoretical effort, the microscopic
mechanism of the photomagnetic effect and the nature of magnetic
ordering remain unclear, often because conclusions are drawn solely on
the basis of macroscopic magnetization measurements
\cite{Pejakovic01SM}.

In this paper we present the results of muon spin relaxation (\musr)
studies in three related Co-Fe PBA compounds: \Cofull\ which is the
stoichiometric limit $x=0.5$, as well as \Kfull\ and \Rbfull.  For
clarity, we abbreviate these henceforth as \Co, \K\ and \Rb\
respectively. Both \K\ and \Rb\ exhibit photomagnetism. In particular,
\K\ shows an enhancement of its ferrimagnetic transition temperature
and magnetization \cite{Sato96S,Pejakovic00PRL}, while \Rb\ shows a
transition from paramagnetic to ferrimagnetic behavior
\cite{Sato99IC}. The results presented here deal with the
unilluminated state these materials, i.e. in absence of light. We find
that even in this state there is some degree of magnetic ordering,
depending on the concentration of vacancies. In both \Co\ and \K\ we
find evidence for static magnetic order. In contrast, the moments
remain dynamic in \Rb, but there is clear indication of magnetic
cluster formation. These results are consistent with a site
percolation model \cite{Stauffer91}, where long range order is
achieved when the concentration of magnetic centers (Fe$^{III}$ and
Co$^{II}$) is above a 3D critical value $p_c=0.3116$
\cite{Stauffer91,Paniagua97EJP}.

\subsection{Experimental} \label{Experimental} 
The three Co-Fe PBA compounds were obtained by aqueous precipitation
followed by high speed centrifugation and drying, and consist of very
fine (sub-micron sized) powders. Several hundred mg of each material
was used in the $\mu$SR experiments. The samples were placed on a
transparent Lucite sample holder and mounted in a horizontal helium
gas flow cryostat with a bore of about $5$~cm coaxial with the muon
beam. Surface muons ($4.1$ MeV) entered the cryostat via thin Kapton
windows. The range of surface muons is about $120$ mg/cm$^2$ (which is
somewhat reduced by the intervening Kapton windows and thin muon
counter). This means that for materials of typical solid densities,
the muons penetrates on the order of $100$ $\mu$m. In addition, the
amount of sample used in our measurements was more than $\sim 200$
mg/cm$^2$ to prevent muons from penetrating through the sample.

The $\mu$SR experiments were performed on the M20 beamline at TRIUMF,
where $100 \%$ spin polarized positive muons (gyromagnetic ratio
$\gamma=13.55$ MHz/kG) are implanted into the sample. The time
evolution of the muon spin polarization depends on the distribution of
internal magnetic fields and their temporal fluctuations. The
implanted muons decay ($\beta^+$ decay with lifetime
$\tau=2.2$~$\mu$s) emitting a positron preferentially along the
direction of the muon spin at the time of decay. The muon spin
polarization as a function of time is thus proportional to the
asymmetry of $\beta$ decay along the initial spin direction. In
magnetic materials, the unpaired spins fluctuate rapidly in the
paramagnetic state, but as the transition temperature is approached
upon cooling, the critical slowing down of the electronic moments
brings the magnetic fluctuations into a time range where they cause
the muon spin to relax. In the ordered state, the moments yield a
static internal magnetic field which affects the precession of the
muon spin. An extensive review of $\mu$SR in magnetic materials is
given in Ref.~\cite{Kalvius01}. Further details on the $\mu$SR
technique may be found in
Refs.~\cite{Kilcoyne98,Chow96PRB,Chow98,Jestadt99PRB}.

For the optical excitation experiments, light was introduced from the
downstream end of the cryostat, with the light source approximately
$1$ m from the sample. The white light intensity from the
tungsten-halogen source used was estimated to be at least $50$
mW/cm$^2$ at the sample, and we used both white and red filtered light
using low pass colored glass (RG665 or OG590) filters from
Melles-Griot. Initially light was introduced via a UVT Lucite
lightguide, but due to concern over IR absorption, this was modified,
first to a $\sim 6$ mm thick Lucite window, then to an IR transparent
Pyrex window. The optical transmission for visible light, in both UVT
Lucite (all thicknesses) and Pyrex ($5$~mm thick window), is better
than $90\%$, ensuring that the samples are illuminated with sufficient
intensity in the range $\sim 550-750$ nm.  In these experiments the
samples were illuminated from the back through the transparent sample
holder and/or from the front (facing the muon beam) using a spherical
mirror with an on-axis aperture to allow the muon beam to arrive at
the sample.

\subsection{Results in \Co: the Stoichiometric Limit $x=0.5$} 
\begin{figure}[h]
\includegraphics[width=\columnwidth]{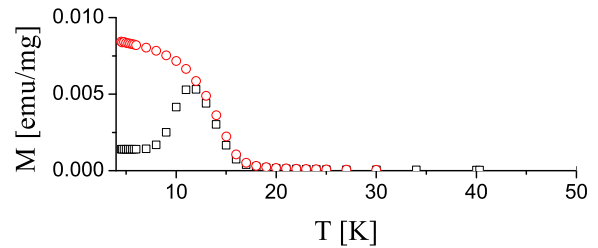}
\caption{The zero field (squares) and field (circles) cooled DC
 magnetization of \Co\ measured at $100$~G in a SQUID magnetometer.}
  \label{MagCo1p5}
\end{figure}

The compound \Co\ was prepared following the procedure described in
Ref.\cite{Sato99IC}. The color, temperature and field dependencies of
the magnetization and the infrared (IR) frequencies of the CN stretch
modes are in agreement with previous work \cite{Sato99IC}. The
magnetization ($M$) was measured as a function of temperature using a
SQUID magnetometer (see Fig.~\ref{MagCo1p5}). Upon zero field cooling
(ZFC), $M$ increases dramatically at $T_c \sim 15$~K and decreases
when cooled further; similarly, the field cooled (FC) magnetization
exhibits an increase at $T_c$ then saturates at low temperatures.
Earlier studies \cite{Sato99IC} indicate that the \Co\ compound
becomes ferrimagnetic at $T_c$ in agreement with our measurements.

\begin{figure}[hbt]
\includegraphics[width=\columnwidth]{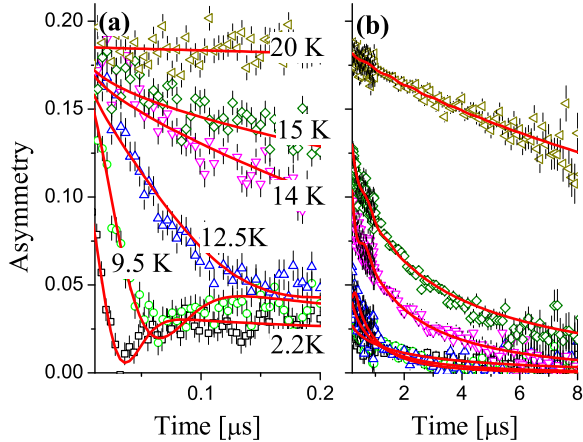}
\caption{(Color online) The muon spin relaxation as a function of time
  measured in \Co\ at $100$~G longitudinal field for different
  temperatures at (a) early and (b) long times. The solid lines are
  fits to the function described in the text.}
 \label{AsyCo1p5}
\end{figure}

The longitudinal field (LF) \musr\ measurements in \Co\ give
information on the behavior of muons in the PBAs structure with dense
Fe$^{III}$ and Co$^{II}$ moments, which serve as a point of comparison
for the other materials studied in this work.  Fig.~\ref{AsyCo1p5}
shows an example of the muon spin relaxation spectra for different
temperatures. Note the asymmetry below $12.5$~K exhibits a dip at
early times $t \sim 0.05$~$\mu$s (Fig.~\ref{AsyCo1p5}(a)), but then
recovers to a higher value at longer times and continues relaxing
slowly down to zero (Fig.~\ref{AsyCo1p5}(b)). This type of muon spin
relaxation is a clear indication of a broad distribution of {\em
  static} local fields at the muon site \cite{Uemura98}.
When the muon experiences a distribution of static magnetic fields
$\rho \left( \frac{\gamma^2({\bf B}-{\bf B}_s)^2}{\Delta^2} \right)$,
where ${\bf B}_s$ is the average static field and $\Delta$ is the root
mean square of the field distribution, the asymmetry follows a static
Kubo-Toyabe function
\begin{eqnarray} \label{KT}
A_{\rm KT}(t)&=&A_0\int \rho \left( \frac{\gamma^2({\bf B}-{\bf
 B}_s)^2}{\Delta^2} \right) G_z(t) d^3B\nonumber \\ 
G_z(t)&=&Re \left\{ \cos^2 \theta + \sin^2 \theta e^{i\gamma B t} \right\}
\end{eqnarray}
where $\theta$ is the angle between the initial muon spin and the
local static magnetic field ${\bf B}$ which is averaged over a powder
sample. For example, if ${\bf B}_s=0$ then the asymmetry is at its
maximum value at $t=0$, it exhibits a dip at $t \sim 1/\Delta$ and
recovers to $\sim 1/3$ its initial value at long times. Depending on
the form of the field distribution, e.g. Gaussian or Lorentzian, the
relaxation follows a Gaussian Kubo-Toyabe (GKT), $A_{\rm GKT}$, or a
Lorentzian Kubo-Toyabe (LKT), $A_{\rm LKT}$, respectively. However, if
in addition to the static field component a small fluctuating field
$B_d(t)$ is present, then the $1/3$ tail continue to relax to zero
\cite{Uemura98}. The relaxation can be described by a phenomenological
function: a LKT or GKT multiplied by a suitable dynamic relaxation
function. The asymmetry in Fig.~\ref{AsyCo1p5} was found to fit best
to GKT multiplied by a square root exponential relaxation,
\begin{equation} \label{AsyGTK}
A(t)=A_{\rm GKT}(t) e^{- \sqrt{\lambda t}}.
\end{equation}
where $\lambda$ is the relaxation rate.

\begin{figure}[htb]
\includegraphics[width=\columnwidth]{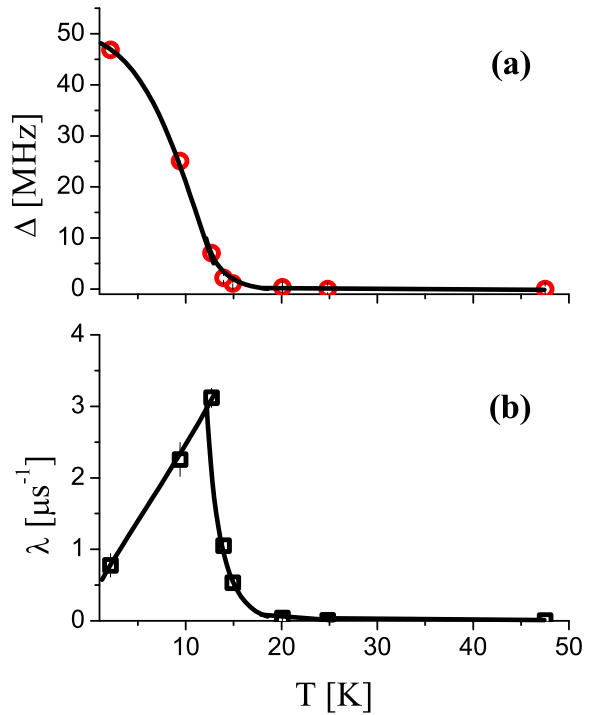}
\caption{The distribution of static fields (a) and the muon spin
   relaxation rate (b) in \Co\ as a function of temperature in an
   applied longitudinal field of $100$~G. The solid lines are a guide
   for the eye.} \label{musrCo1p5}
\end{figure}

The static field distribution width $\Delta$ obtained from the fits is
presented in Fig.~\ref{musrCo1p5}(a). In the paramagnetic state above
$15$~K, $\Delta=0$, as expected from a fully dynamic field at the muon
site. Below $15$~K, $\Delta$ increases dramatically as the magnetic
moments of Co$^{II}$ and Fe$^{III}$ freeze, generating a static field
distribution. The size of this static field increases to $\Delta \sim
47$~MHz, corresponding to a width in field of $\sim 3.5$~kG at low
temperatures. The dynamic relaxation rate, $\lambda$, exhibits a sharp
increase below $20$~K, peaks at $\sim 12.5$~K and decreases slowly
upon further cooling (Fig.~\ref{musrCo1p5}(b)). The temperature
dependence of $\lambda$ is indicative of a sharp magnetic ordering
phase transition at $T_c \sim 12.5$~K, where the increase as $T$
approaches $T_c$ from above is due to critical slowing down of the
fluctuations.  The relaxation at low temperatures is attributed to
remnant local field fluctuations due to spin wave excitations
\cite{Kalvius01}. Both $\lambda$ and $\Delta$ temperature dependencies
are in agreement with the ferrimagnetic behavior seen in the
magnetization measurements.

Note that we find no missing fraction in low transverse fields in the
paramagnetic phase, indicating no appreciable Muonium\footnote{Muonium
  (a $\mu^+ e^-$ bound state) is known to react with CN
  \cite{Stadlbauer83JPC}.} formation. The absence of spontaneous spin
precession in the magnetic phase in this nominally stoichiometric
Co-Fe PBA and the square root relaxation behavior indicates that even
here disorder, size and shape distribution of grain size and/or
multiple inequivalent muon sites are sufficient to give a broad field
distribution. In contrast, if $\gamma B_s \gg \Delta$ one expects muon
spin precession at frequency $\gamma B_s$. The magnitude of the
typical field (measured by $\Delta$) also gives an indication of the
size of the fields that can be expected in the photomagnetic
compositions, as well as their temperature dependencies. This behavior
will be compared to that observed in the photomagnets \K\ and \Rb.

\subsection{Results in \K: $x=0.4$}
We followed the procedure described in Ref.~\cite{Sato96S} to prepare
the \K\ compound. Both the color and magnetization are in agreement
with previous work \cite{Sato96S}. \K\ is an example where red light
changes the magnetization and the transition temperature
\cite{Sato96S,Pejakovic01SM}.
\begin{figure}[htb]
\includegraphics[width=\columnwidth,clip]{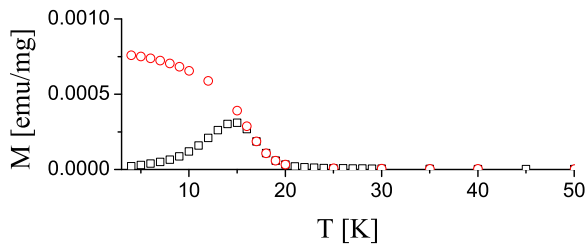}
\caption{The zero field (squares) and field (circles) cooled DC
 magnetization of \K\ measured at $100$~G in a SQUID magnetometer.}
 \label{MagK}
\end{figure}
Similar to the \Co, \K\ undergoes magnetic ordering below $20$~K even
in the unilluminated state. This is evident in the magnetization of
\K\ which was measured in a SQUID magnetometer (see Fig.~\ref{MagK}),
and shows a dramatic increase at $T_c \sim 15$~K in both the FC and
ZFC magnetization, in agreement with Ref.~\cite{Sato96S}.

\begin{figure}[htb]
\includegraphics[width=\columnwidth]{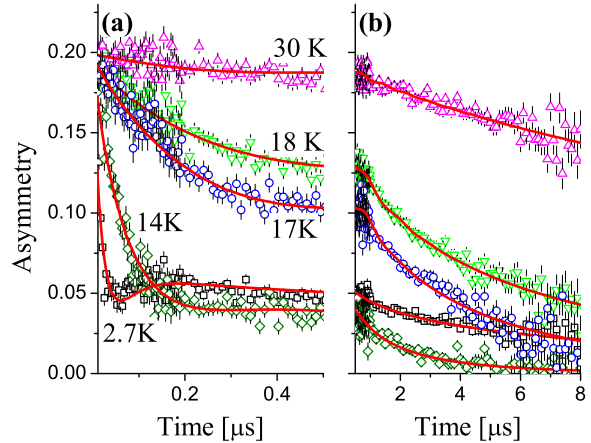}
\caption{(Color online) The muon spin relaxation as a function of time
  measured in \K\ at $100$~G longitudinal field for different
  temperatures at (a) early and (b) long times. The solid lines are
  fits to the function described in the text}
 \label{AsyK}
\end{figure}
The muon relaxation in \K\ is similar to that found in \Co\ (see
Fig.~\ref{AsyK}), where at low temperatures the muons experience a
local field which is a combination of a static distribution of local
fields and an additional small fluctuating component. However, the
relaxation in this case is better described by a LKT multiplied by a
square root exponential function,
\begin{equation}
A(t)=A_{\rm LKT}(t) e^{-\sqrt{\lambda t}}.
\end{equation}
A Lorentzian static field distribution is typical in dilute spin glass
systems where the breadth of the distribution is due to the
distribution of muon sites with varying distances from the magnetic
moments \cite{Uemura98}.

\begin{figure}[htb]
\includegraphics[width=\columnwidth]{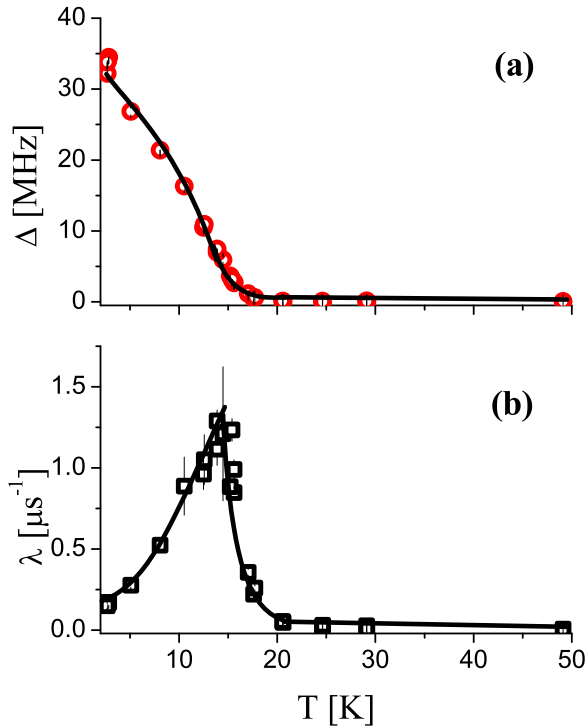}
\caption{The distribution of static fields (a) and muon spin
  relaxation rate (b) in \K\ as a function of temperature in a
  longitudinal field of $100$~G. The solid lines are a guide for the
  eye.}
\label{musrK}
\end{figure}

The temperature dependencies of the fit parameters $\Delta$ and
$\lambda$ in \K\ are similar to those seen in \Co. The distribution of
static fields $\Delta$, presented in Fig.~\ref{musrK}(a), is zero
above $15$~K, indicating the muon spin relaxation is due entirely to a
fluctuating field. Below this temperature $\Delta$ increases and
saturates at low temperatures, where $\Delta=33$~MHz corresponding to
a width in field of $\sim 2.5$~kG, slightly smaller than that found in
\Co, as expected from a system with lower moment concentration.
The relaxation rate $\lambda$, presented in Fig.~\ref{musrK}(b), has a
sharp increase below $20$~K, peaks at $T \sim 12.5$~K, indicative of
the slowing down of magnetic field fluctuations, and decreases slowly
at lower temperatures. As expected the relaxation rate in \K\ is also
lower than in \Co, consistent with the smaller average local field
experienced by muons in \K.

\subsection{Results in \Rb: $x \simeq 0.17$}
\begin{figure}[htb]
\includegraphics[width=\columnwidth]{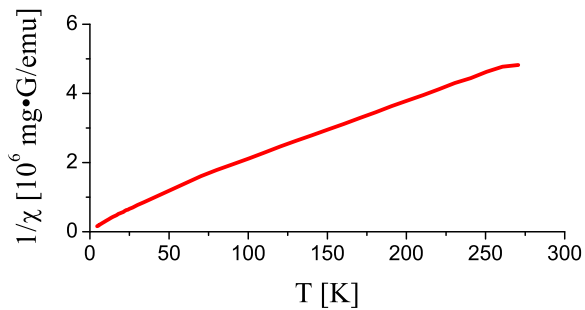}
\caption{The inverse susceptibility of \Rb\ as a function of
  temperature measured at $100$~G.}
\label{MagRb}
\end{figure}

The compound \Rb\ was prepared according to the procedure described in
Ref.\cite{Sato99IC}. The color, magnetization and the IR frequencies
of CN stretching mode of the compound are in agreement with previous
work \cite{Sato99IC}. The magnetization as a function of temperature
at $100$~G is featureless and very small compared to that in \Co\ and
\K\ due to the high concentration of diamagnetic centers
Fe$^{II}(t_{2g}^6,S=0)$-CN-Co$^{III}(t_{2g}^6,S=0)$. As can be seen in
Fig.~\ref{MagRb}, the inverse susceptibility ($1/\chi$) exhibits an
almost linear\footnote{Note that the value of the magnetization for
this paramagnetic compound is rather low, therefore the small
deviation from the perfect linear behavior may be due to small
inaccuracies in the measurements.} dependence on temperature, indicating
that unilluminated \Rb\ remains paramagnetic down to at least $4$~K.
In contrast to \Co, paramagnetic \Rb, is expected to show a strong
response to visible light \cite{Sato99IC,Goujon01P}. 

\begin{figure}[htb]
\includegraphics[width=\columnwidth]{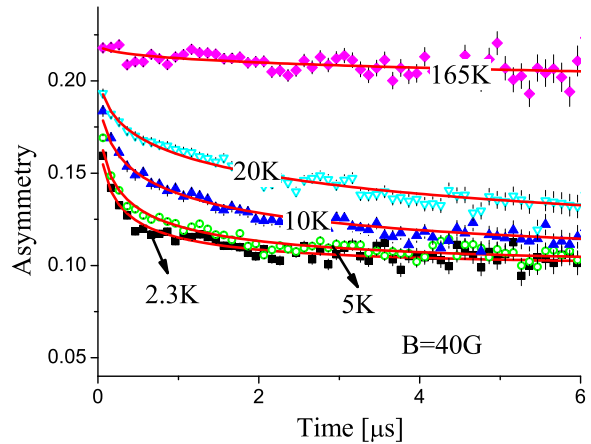}
\caption{(Color online) The muon spin relaxation in \Rb\ at LF
  $B=40$~G and different temperatures.}
\label{SpectraT40G}
\end{figure}

\begin{figure}[htb]
\includegraphics[width=\columnwidth]{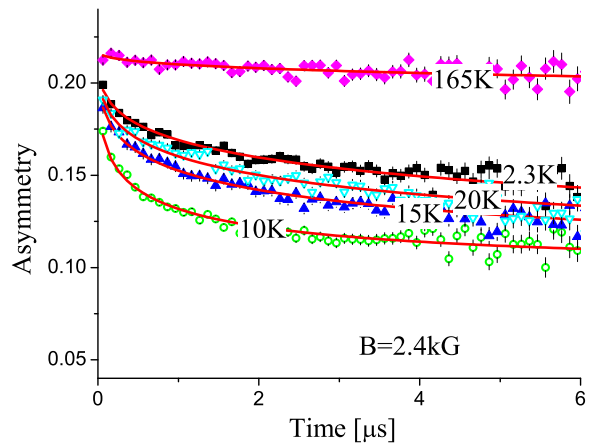}
\caption{(Color online) The muon spin relaxation in \Rb\ at $B=2.4$~kG
  and different temperatures.}
\label{SpectraT2p4kG}
\end{figure}

The muon spin relaxation was measured in non-stoichiometric \Rb\ at
different longitudinal fields and temperatures between $300$~K and
$2.3$~K. The asymmetry at low LF ($B=40$~G) and various temperatures
is shown in Fig.~\ref{SpectraT40G}. At this field the relaxation is
slow at high temperatures and increases monotonically at low
temperatures.
In contrast, at high LF ($B=2.4$~kG) the relaxation increases as the
temperature is decreased, peaks at $T \sim 10$~K and then decreases at
lower temperatures (see Fig.~\ref{SpectraT2p4kG}).
Note the asymmetry measured in \Rb\ shows no dip/recovery at early
times, indicating that the muons do not experience a static field
component in this compound, and that the spin relaxation here is
entirely dynamic in origin.

All spectra were fit with the sum of a stretched exponential and a
non-relaxing background,
\begin{equation} \label{DynamicRlx}
A(t)=A_0 \exp{(-(\lambda t)^{0.35})}+Bg.
\end{equation}
The background asymmetry, $Bg=0.1$, is quite large and is due to muons
missing the sample ($9$~mm diameter) and stopping in the silver mask
around it. In contrast, the other samples were large ($15$~mm
diameter) and had little or no background, as is evident in
Fig.~\ref{AsyCo1p5}(b) and \ref{AsyK}(b). The relaxation rate
$\lambda$ is the average spin lattice relaxation (SLR) rate. The small
value of the exponent $0.35$ characterizing the distribution of
relaxation rates is typical in dilute spin glass systems
\cite{Uemura98}, suggesting a similar distribution of magnetic moments
in \Rb.

\begin{figure}[htb]
\includegraphics[width=\columnwidth]{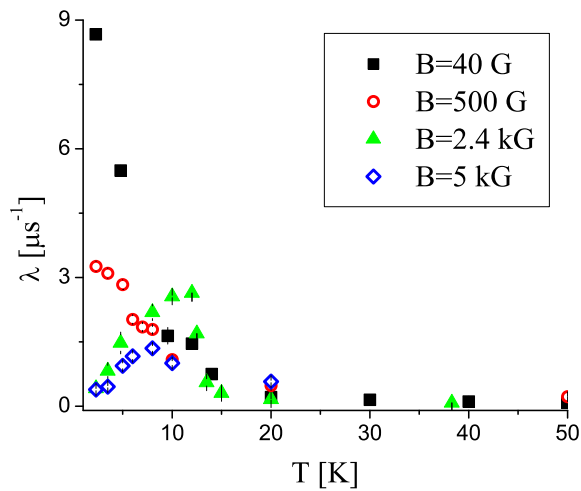}
\caption{(Color online) The muon spin relaxation rate in \Rb\ as a
  function of temperature at different longitudinal fields.}
\label{musrRb}
\end{figure}

The average relaxation rate for all temperatures and fields
investigated is presented in Fig.~\ref{musrRb}. Above $20$~K the
relaxation rate is field independent. However, below this temperature
$\lambda$ depends strongly on field and increases significantly as the
temperature is lowered.
At low fields ($40-1000$~G) the relaxation rate does not exhibit a
peak, while at fields greater than $2.4$~kG the relaxation rate peaks
and then decreases at lower temperatures. This type of behavior is
characteristic of magnetic cluster freezing seen in other molecular
magnetic materials
\cite{Salman02PRB,Salman00PB,Jestadt01JPCM,Lascialfari98PRL,Blundell03P,Blundell04JPCM}.
Molecular magnets consist of clusters of magnetic ions with a large
exchange coupling $J$ between them, while a very weak dipolar coupling
exists between neighboring molecules, such that at low temperatures
they behave as non-interacting large magnetic moments (${\bf m}$). In
these systems a large increase in the relaxation rate is observed at
low LF (${\bf m} \cdot {\bf B} \ll J$) when the temperature is
comparable to $J$. This increase is due to slowing down of the local
field fluctuations at the muon site as the clusters coalesce. However,
at high LF (${\bf m} \cdot {\bf B} \gg J$) and low temperature the
relaxation rate decreases as fluctuations in the molecular moments are
quenched by the applied field, i.e. as the thermal energy becomes
smaller than the molecular Zeeman splitting $k_B T < {\bf m} \cdot
{\bf B}$.

These results strongly disagree with the conclusion, based on
magnetization measurements similar to those of Fig.\ref{MagRb}, that
\Rb\ remains a simple paramagnet down to low temperature
\cite{Sato99IC}. Our $\mu$SR measurements show that even in the
unilluminated ground state, a disordered magnetic state is forming
below $20$~K that is not apparent in the macroscopic magnetization.

The time-scale and size of the dynamics of the local field at the muon
site can be estimated from the dependence of the muons' SLR on the
applied LF at low temperature. In the fast fluctuation limit the SLR
time follows \cite{Uemura85PRB,Keren94PRB,Salman02PRB}
\begin{subequations}
\label{T1vsH}
\begin{eqnarray}
T_{1}(B) &=&\alpha+\beta B^{2},  \label{T1vsHa} \\
\alpha &=&\frac{1}{\Delta ^{2}\tau },  \label{T1vsHb} \\
\beta &=&\frac{(2 \pi \gamma)^{2}\tau }{\Delta ^{2}}. \label{T1vsHc}
\end{eqnarray}
\end{subequations}
The correlation time ($\tau$) and mean square of the transverse field
distribution at the muon site in frequency units ($\Delta^{2}$) are
defined by the auto-correlation function of the local transverse
field,
\begin{equation} 
\gamma ^{2}\left\langle {\bf B}_{\bot }(t){\bf B}_{\bot}(0)\right\rangle
=\Delta ^{2}\exp \left( -t/\tau \right) . \label{Bcorrelation}
\end{equation}
The fast fluctuation limit is defined by $\tau \Delta < 1$
\cite{Keren94TH}. Note that in Eq.\ref{T1vsH} and \ref{Bcorrelation}
an exponential spin relaxation function is assumed. However, the muons
can occupy magnetically inequivalent sites in the lattice, and
therefore experience a distribution of $\Delta$ values. As a result
the spin relaxation becomes non-exponential
(e.g. Eq.~\ref{DynamicRlx}) and one should average over all possible
values of $\Delta$ \cite{Salman02PRB}. The linear relation in
Eq.~\ref{T1vsH} still holds in this case, but $\Delta$ is interpreted
as the average value of the local field distribution widths for all
possible sites.

\begin{figure}[htb]
\includegraphics[width=\columnwidth]{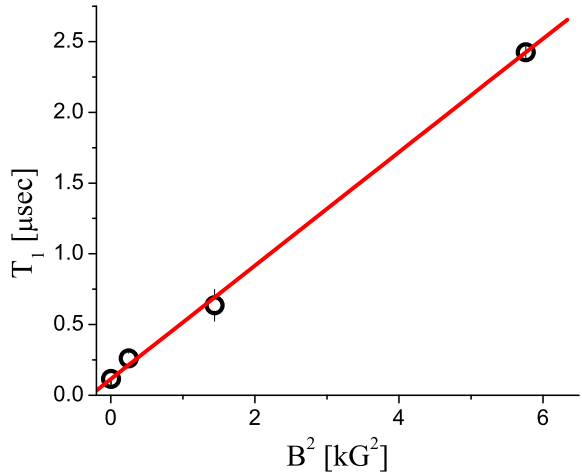}
\caption{The muon spin relaxation time at $T=2.3$~K as a function of
the magnetic field squared. The solid line is a linear fit $T_1=\alpha
+\beta B^2$ with $\alpha=0.1147(9)$~$\mu$s and
$\beta=0.401(1)$~$\mu$s/kG$^2$.}
\label{T1vsH2}
\end{figure}

In Fig.~\ref{T1vsH2} we plot the SLR time, $T_1 \equiv 1/\lambda$, at
$T=2.3$~K as a function of the applied magnetic field squared, $B^2$.
We find that, as expected from Eq.~\ref{T1vsH}, the relaxation time is
proportional to $B^2$. The correlation time $\tau$ and the size of the
local magnetic field calculated from the free constant
$\alpha=0.1147(9)$~$\mu$s and the slope $\beta=0.401(1)$~$\mu$s/kG$^2$
of the linear relation \cite{Salman02PRB}, are $\Delta=19.93(5)$~MHz
and $\tau=21.9(1)$~ns, justifying the assumption that we are in the
fast fluctuation limit. $\Delta$ corresponds to a width in field of
$0.587(2)$~kG, much smaller than that measured in \Co\ and \K, and
consistent with the smaller concentration of magnetic moments in \Rb.

\subsection{Sample Illumination}
We attempted to observe photoinduced changes in the magnetism of \K\
and \Rb\ in a range of different configurations of sample mounting and
optical illumination as discussed in section \ref{Experimental}.  In
other measurements
\cite{Sato96S,Sato99IC,Pejakovic01SM,Pejakovic00PRL}, similar light
intensity was found to be sufficient to observe the photoinduced
effect in both compounds. However, in neither back nor front
illumination configurations, even for illumination times up to $\sim
5$ hours at $T=5$~K, did we observe a significant effect of the light
on the $\mu$SR signals in \K\ or \Rb.  Heating effects of illumination
were apparent in the sample thermometry, but were at most a few
degrees.  This is still well below the thermal reversion of the
metastable state above $\sim 100$ K \cite{Sato99IC}.

The lack of a photomagnetic effect in our data is probably the result
of a mismatch between the stopping range of muons and the absorption
length of light. Co-Fe PBAs are highly colored materials, but are also
very opaque, a consequence of low energy $d$-$d$ and metal-ligand
charge transfer excitations. It has been shown that with the formation
of the metastable magnetic centers, the material becomes less
absorbing in the wavelength region relevant to the photomagnetism
\cite{Sato02BCSJ}, thus it is possible that the photomagnetic state
would eventually grow out from the illuminated surfaces, but such an
effect was apparently not sufficient to influence the magnetism on the
$\sim 100$ $\mu$m surface muon stopping range. A more favorable
situation to study these effects is in a thin film geometry
\cite{Park04JMMM} using low energy muons \cite{Bakule04CP} or
$\beta$-NMR probes \cite{Morris04PRL,Salman04PRB,Salman06PRL}.

\subsection{Summary and Conclusions}
The close similarity of the temperature dependencies of $\Delta$ and
$\lambda$ in \Co\ and \K\ are strong indications that both compounds
undergo similar magnetic transitions, from dynamic paramagnetism to
long range magnetic order. However, in \Rb\ there is no evidence of
freezing of the magnetic moments. We now present these results within
the context of a simple site percolation model
\cite{Stauffer91}. Assuming that the fraction of magnetic (Fe$^{III}$
and Co$^{II}$) to non-magnetic (vacancies) centers in the system is
$p$, the percolation theory predicts that above a critical value
$p_c=0.3116$ \cite{Stauffer91,Paniagua97EJP} magnetic long range order
is possible. In our compounds the number of non-magnetic centers is
$1-2x$, therefore $p \equiv 2x$. In particular $p=1,0.8$ and $\sim
0.34$ for \Co, \K\ and \Rb, respectively. Note that $p$ is well above
the critical value for both \Co\ and \K, and consequently long range
magnetic order is expected, while for the \Rb\ compound $p$ is close
to the critical value, and therefore the system may not be able to
achieve such order. Indeed, our results are consistent with long range
magnetic order formed at low temperature in \Co\ and \K, but not in
\Rb.

\begin{table}[h]
\begin{ruledtabular}
\begin{tabular}{|l|c|c|c|}
Compound & $p$   & $\Delta$         & $\Delta$      \\ 
         &       & [MHz]          & [kG]            \\ \hline
 \Co     & $1.0$ & $46.8 \pm 1.2$   & $3.45 \pm 0.09$   \\
 \K      & $0.8$ & $33.3 \pm 1.0$   & $2.46 \pm 0.07$   \\
 \Rb     &$0.34$ & $19.93 \pm 0.05$ & $0.587 \pm 0.002$ \\ 
\end{tabular}
\end{ruledtabular}
\caption{Summary of the average value of local field at the muon
  site, measured at $T \sim 2.3$~K.} \label{Del}
\end{table}
The size of local magnetic field at low temperature is summarized in
Table~\ref{Del}. Although estimated from different muon relaxation
behavior in the three compounds, $\Delta$ can be used as an estimate
of the size of the local field, which clearly decreases with
decreasing $p$ as expected for a lower concentration of magnetic
centers and a smaller average magnetic cluster size.

As Figs.~\ref{musrCo1p5}, \ref{musrK}, and \ref{musrRb} demonstrate,
the common energy scale that appears in all three compounds is $T\sim
12.5$~K. This corresponds to the measured strength of the
superexchange coupling $J=15$~K between neighboring Fe$^{III}$ and
Co$^{II}$ in \Co\ \cite{Sato99IC}. Since the chemical structure and
the CN bond length away from vacancies is identical in \Co, \K\ and
\Rb, the coupling between Fe$^{III}$ and Co$^{II}$ pairs is expected
to be very similar in all three compounds. Therefore, in the
unilluminated state $J$ is independent of the alkali metal M
concentration $1-p$. Note however, that this is not necessarily true
near vacancies, and therefore after illumination the coupling between
Fe$^{III}$ and Co$^{II}$ pairs may be different.

These results indicate that the magnetic properties of the Co-Fe PBAs
in the unilluminated state depend strongly on the concentration of
vacancies. We find evidence for long range order when the
concentration of magnetic centers is above the critical value, $p_c$,
as is the case in \Co\ and \K, while near $p_c$ we find evidence of
magnetic cluster formation without freezing, in agreement with a
simple site percolation model.

\begin{acknowledgments}
  We thank N.D.~Draper and D.B.~Leznoff of Simon Fraser University for
  advice with the sample preparations, R.F.~Kiefl for useful
  discussions, and B.~Hitti and R.~Abasalti for technical assistance
  with the measurements. This work was funded by TRIUMF and NSERC of
  Canada.
\end{acknowledgments}

\newcommand{\noopsort}[1]{} \newcommand{\printfirst}[2]{#1}
  \newcommand{\singleletter}[1]{#1} \newcommand{\switchargs}[2]{#2#1}

\end{document}